\documentclass[conference]{IEEEtran}
\usepackage{graphicx,psfrag,epsfig,epsf,latexsym,hhline,amsmath,amssymb,multirow}
\usepackage[usenames,dvipsnames]{pstricks}
\usepackage{pst-plot}
\usepackage{pstricks-add}
\usepackage{graphicx}
\usepackage{amsthm}
\usepackage{algorithm}
\usepackage{algpseudocode}
\usepackage[noadjust]{cite}
\usepackage{blindtext}
\usepackage{etoolbox}
\IEEEoverridecommandlockouts
\graphicspath{ {figures/} }

\input{Jerry.def}

\begin{document}
\title{Golden-Coded Index Coding}
\author{\dag Yu-Chih Huang, \ddag Yi Hong, and \ddag Emanuele Viterbo\\
\dag Department of Communication Engineering, National Taipei University\\
\ddag Department of Electrical and Computer Systems Engineering, Monash University\\
{\tt\small {\{ychuang@mail.ntpu.edu.tw, yi.hong@monash.edu, emanuele.viterbo@monash.edu\}} }
\thanks{The work of Y.-C. Huang was supported by Ministry of Science and Technology, Taiwan, under grant MOST 104-2218-E-305-001-MY2. The work of Yi Hong and Emanuele Viterbo is supported by the Australian Research Council (ARC) through the Discovery Project under grant DP160101077.}}

\maketitle

\begin{abstract}
    We study the problem of constructing good space-time codes for broadcasting $K$ independent messages over a MIMO network to $L$ users, where each user demands all the messages and already has a subset of messages as side information. As a first attempt, we consider the $2\times 2$ case and propose \textit{golden-coded index coding} by partitioning the golden codes into $K$ subcodes, one for each message. The proposed scheme is shown to have the property that for any side information configuration, the minimum determinant of the code increases \textit{exponentially} with the amount of information contained in the side information.
\end{abstract}


\section{Introduction}
As the recent rise of wireless caching and cache-enabled cloud RAN for 5G systems \cite{niesen14,ji16,paschos16}, it is more and more likely that one will face the scenario where one or multiple senders wish to broadcast to multiple receivers which already have some messages as side information. Depending on the application, side information could be pre-stored contents at the receivers during off-peak hours or could be packets decoded from the previous sessions. At the network layer, this problem is called index coding \cite{bar-yossef11} and has been studied intensively; however, joint design of physical-layer coding/modulation and index coding is relatively less investigated.

In this work, we study a particular case where the receivers demand \textit{all the messages}, i.e., multicasting. For this case, the problem has been previously studied for the AWGN channel \cite{viterbo14index}, where a new class of codes named lattice index codes based on lattice codes is proposed to mimic the behavior of capacity-achieving codes. The lattice index codes in \cite{viterbo14index} are shown to have the minimum squared Euclidean distance increasing exponentially as the rate of side information for any side information configuration. Moreover, when normalized by the rate of side information, the SNR difference between the codes with and without side information for achieving the same error probability is $6$ dB/bit. This property is called uniform side information gain.

In \cite{huang15_lic}, the same problem was studied for the Rayleigh fading channel in which the minimum product distance is much more important than the minimum Euclidean distance. The new lattice index code construction for the Rayleigh fading channel in \cite{huang15_lic} provides exponentially increased squared minimum product distance as the rate of side information increases and provides uniform side information gain for any side information configuration.

In this paper, we turn our focus to the scenario that is frequently seen in almost every modern wireless communication systems, the multiple-input multiple-output (MIMO) fading channel, to accommodate multiple antennas. We first analyze the probability of error and derive an approximation of SNR gain provided by side information as a function of the minimum squared determinants and the numbers of codewords having the minimum determinant of the codebooks with and without side information.

We then study the construction of good space-time index codes. While there is a rich literature in the study of construction of space-time codes for the point-to-point MIMO channel (see \cite{oggier07} and the reference therein), as a first attempt, we consider construction of lattice space-time index codes solely based on golden codes \cite{belfiore05GC} for the $2\times 2$ case. The main difficulty is that most of the code constructions proposed in \cite{viterbo14index} and \cite{huang15_lic} rely on partitions induced by the Chinese remainder theorem (CRT) for some commutative rings; however, golden codes (and most of the lattice space-time codes) are constructed over a cyclic division algebra, which is non-commutative and hence prevents the direct application of CRT. We overcome this challenge and propose the \textit{golden-coded index coding} by making connection between the underlying cyclic division algebra and a ring of algebraic integers and then partitioning this ring instead. The proposed golden-coded index coding is shown to provide minimum determinant, which exponentially increases as the rate of side information increases and uniform side information gain of 6 dB for any side information configuration. We also use simulations to verify the theoretic analysis and show that the approximation derived in this paper can accurately predict the actual side information gain.

The rest of the paper is organized as follows. In Section~\ref{sec:prob_statement}, we provide a formal description of the problem of broadcasting over a MIMO channel with message side information at receivers. We then partition the maximal order of the golden algebra, a cyclic division algebra over which the golden code is constructed and propose golden-coded index coding in Section~\ref{sec:GC_IC}. Simulation results are given in Section~\ref{sec:simu_result} to verify the validity of the analysis in this paper and some concluding remarks are given in Section~\ref{sec:conclude}.

\section{Problem Statement}\label{sec:prob_statement}
\begin{figure}
    \centering
    \includegraphics[width=2.5in]{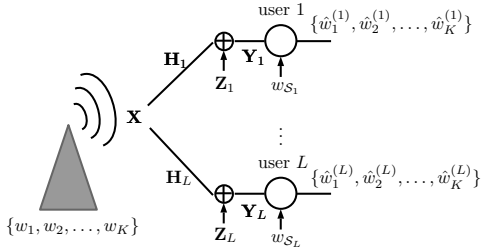}
    \caption{Broadcast over MIMO channel with message side information.}
    \label{fig:system_model}
\end{figure}
We consider a network with a base station equipped with $n_t$ antennas and $L$ users each equipped with $n_r$ antennas as shown in Fig~\ref{fig:system_model}. The base station broadcasts $K$ independent messages $\{w_1,\ldots,w_K\}$ with $w_k$ uniformly distributed over $\{1,\ldots,W_k\}$ to the $L$ users, where each user $l$ demands all the messages and already has a subset of messages $w_{\mc{S}_l}\defeq \{w_k|k\in\mc{S}_l\}$ governed by the index set $\mc{S}_{l}\subseteq \{1,\ldots,K\}$ as side information. The signal emitted from the base station is spread over $T$ symbol durations and can be represented as a $n_t\times T$ matrix $\mathbf{X}$ where each entry is subject to the power constraint $\mbb{E}[|x_{jt}|^2]=1$. The signal received at the $l$-th user can be represented as a $n_r\times T$ matrix given by
\begin{equation}
    \mathbf{Y}_l = \mathbf{H}_l\mathbf{X} + \mathbf{Z}_l,
\end{equation}
where $\mathbf{H}_l$ is a random $n_r\times n_t$ matrix with each element i.i.d. distributed $\mc{CN}(0,1)$ and $\mathbf{Z}_l$ is a random $n_r\times T$ matrix with each element i.i.d. distributed $\mc{CN}(0,\sigma^2_l)$. The signal-to-noise ratio (SNR) is then defined as $\mathrm{SNR}_l\defeq \frac{n_t}{\sigma^2_l}$.

We assume perfect channel state information $\mathbf{H}_l$ is available at receiver $l$. After receiving $\mathbf{Y}_l$, the receiver $l$ forms $\{\hat{w}_1^{(l)},\ldots,\hat{w}_K^{(l)}\}$ an estimate of $\{w_1,\ldots,w_K\}$ according to $\mathbf{Y}_l$ and $w_{\mc{S}_l}$. The probability of error at the user $l$ is defined as
\begin{equation}
    p_e^{(l)} \defeq \Pp\{ \{w_1,\ldots,w_K\} \neq \{\hat{w}_1^{(l)},\ldots,\hat{w}_K^{(l)}\}\}.
\end{equation}

Let $\mc{C}$ be the transmitted codebook and the encoder maps the messages to the codewords as $f(w_1,\ldots,w_K)=\mathbf{X}\in \mc{C}$. For any pair of codeword matrices $\mathbf{X},\mathbf{X}'\in\mc{C}$, let $\mathbf{A}\defeq (\mathbf{X}-\mathbf{X}')(\mathbf{X}-\mathbf{X}')^{\dag}$ and let $r$ be the rank of $\mathbf{A}$. For a generic receiver, without any side information, in the high SNR regime, one has an upper bound on $\Pp(\mathbf{X}\rightarrow\mathbf{X}')$ the pairwise error probability as follows \cite{oggier07},
\begin{equation}\label{eqn:pairwise_pe}
    \Pp(\mathbf{X}\rightarrow\mathbf{X}') \leq \left(\frac{\mathrm{SNR}\Delta^{1/r}}{4 n_t}\right)^{-rn_r},
\end{equation}
where $\Delta =\prod_{m=1}^r\lambda_m$ with $\lambda_1,\ldots,\lambda_m$ being the non-zero eigenvalues of $\mathbf{A}$. In this work, we further restrict our attention to full rank codes in which $r=n_t$ and
\begin{equation}
    \Delta =\prod_{m=1}^{n_t}\lambda_m = \det(\mathbf{A})\neq 0.
\end{equation}
One can also define the {\em minimum determinant} of $\mc{C}$ as
\begin{equation}\label{eqn:min_det_C}
    \delta \defeq \min_{\mathbf{X}\neq\mathbf{X'}\in\mc{C}}\det(\mathbf{A}).
\end{equation}
If $\mc{C}$ is carved from a lattice, \eqref{eqn:min_det_C} can be further rewritten as
\begin{equation}\label{eqn:min_det_C_lattice}
    \delta \defeq \min_{\mathbf{X}\neq\mathbf{0}\in\mc{C}} \det(\mathbf{X})^2.
\end{equation}
Let $N_{\mathbf{X}}$ be the number of codewords $\mathbf{X'}\in\mc{C}$ such that the corresponding $\mathbf{A}$ has determinant $\delta$. Also, let
\begin{equation}\label{eqn:N_avg}
    N_{\mc{C}}\defeq \frac{1}{|\mc{C}|}\sum_{\mathbf{X}\in\mc{C}} N_{\mathbf{X}},
\end{equation}
be the average number of codewords having $\delta$ to a codeword in $\mc{C}$. The probability of error of a code carved from a lattice can then be approximated as
\begin{align}\label{eqn:union_bound}
    p_e &= \frac{1}{|\mc{C}|}\sum_{\mathbf{X}\in\mc{C}} \Pp\left(\bigcup_{\mathbf{X}'\neq\mathbf{X}} \mathbf{X}\rightarrow\mathbf{X}'\right) \nonumber \\
    &\overset{(a)}{\approx} \frac{1}{|\mc{C}|} \sum_{\mathbf{X} \in\mc{C}} N_{\mathbf{X}}\left(\frac{\mathrm{SNR} \delta^{1/n_t}}{4 n_t}\right)^{-n_t n_r} \nonumber \\
    &\overset{(b)}{=} N_{\mc{C}}\left(\frac{\mathrm{SNR} \delta^{1/n_t}}{4 n_t}\right)^{-n_t n_r},
\end{align}
where in ($a$) we have applied union bound only to codewords having $\delta$ to $\mathbf{X}$ and ignored all the other terms and ($b$) is from the definition of $N_{\mc{C}}$ in \eqref{eqn:N_avg}.

Now, with the help of side information $w_{\mc{S}_l}$, the receiver $l$ can expurgate all the codewords which do not correspond to $w_{\mc{S}_l}$ and form the subcode $\mc{C}_{\mc{S}_l}\defeq \{ f(v_1,\ldots,v_K)|v_k=w_k,\forall k\in\mc{S}_l\}$.  It is clear that $\mc{C}_{\mc{S}_l}\subseteq \mc{C}$ and $\delta_l$ the minimum determinant associated with $\mc{C}_{\mc{S}_l}$ is no less than $\delta$, i.e., $\delta_l\geq \delta$. It is of primary interest to investigate when $\delta_l > \delta$ and how is this gain translated into the SNR gain. To this end, we let $\mathrm{SNR}_l$ be the SNR required for the codebook $\mc{C}_{\mc{S}_l}$ to achieve the same error probability $p_e$ which can be achieved by using $\mc{C}$ with $\mathrm{SNR}$. Plugging these parameters into \eqref{eqn:union_bound} leads to
\begin{align}
    &N_{\mc{C}}\left(\frac{\mathrm{SNR} \delta^{1/n_t}}{4 n_t}\right)^{-n_t n_r} \approx N_{\mc{C}_{\mc{S}_l}}\left(\frac{\mathrm{SNR}_l \delta_l^{1/n_t}}{4 n_t}\right)^{-n_t n_r} \nonumber \\
    &(\Leftrightarrow)\quad 10\log_{10}(\mathrm{SNR}) - 10\log_{10}(\mathrm{SNR}_l) \approx \nonumber \\
    &\hspace{1.5cm} \frac{1}{n_tn_r}10\log_{10}\left(\frac{N_{\mc{C}}}{N_{\mc{C}_{\mc{S}_l}}}\right) + \frac{1}{n_t}10\log_{10}\left(\frac{\delta_l}{\delta}\right) \nonumber \\
    &(\Leftrightarrow)\quad \text{SNR gain} \text{~of revealing $w_{\mc{S}_l}$ in dB} \approx \nonumber \\
    &\hspace{1.5cm} \frac{1}{n_tn_r}10\log_{10}\left(\frac{N_{\mc{C}}}{N_{\mc{C}_{\mc{S}_l}}}\right) + \frac{1}{n_t}10\log_{10}\left(\frac{\delta_l}{\delta}\right). \label{eqn:snr_gain_exact}
\end{align}
This term provides a fairly accurate estimate on the SNR gain obtained from revealing $w_{\mc{S}_l}$. However, it is in general difficult to control both $N_{\mc{C}_{\mc{S}_l}}$ and $\delta_l$ for lattice codes. Hence, we follow the approach taken by most of the work in the literature (see \cite{oggier07} and reference therein), which only focuses on $\delta_l$ and redefine the SNR gain as $10\log_{10}\left(\delta_l/\delta\right)^{n_t}$ dB (second term in \eqref{eqn:snr_gain_exact}). Moreover, since we wish to understand how the SNR gain scales with the amount of information contained in the side information, we therefore define the side information gain of the code $\mc{C}$ and the index set $\mc{S}_l$ for the MIMO broadcast network as
\begin{equation}\label{eqn:side_info_gain}
    \Gamma(\mc{C},\mc{S}_l)\defeq \frac{10\log_{10}\left(\frac{\delta_l}{\delta}\right)}{n_t R_{\mc{S}_l}},
\end{equation}
where $R_{\mc{S}_l} \defeq \sum_{k\in\mc{S}_l} R_k$ with $R_k$ being the rate (bits per real dimension) of the message $w_k$. This side information gain essentially serves as a (rough) approximation of the SNR gain (in dB/bits) provided by side information $w_{\mc{S}_l}$. We again would like to emphasize that a better approximation is to use the equation in \eqref{eqn:snr_gain_exact}. Throughout the paper, we will use \eqref{eqn:side_info_gain} as the design guideline and use \eqref{eqn:snr_gain_exact} to explain the simulation results.

\section{Proposed Golden-Coded Index Coding}\label{sec:GC_IC}
In this section, we review the golden code for the $2\times 2$ MIMO case and propose golden-coded index coding.

\subsection{Golden algebra and golden codes}
Consider $\mbb{Q}(i,\sqrt{5})$ a quadratic extension of $\mbb{Q}(i)$ and $\sigma:\sqrt{5}\rightarrow -\sqrt{5}$ its non-trivial $\mbb{Q}(i)$-automorphism. The golden code is built from the cyclic division algebra (golden algebra)
\begin{equation}
    \mc{A}=(\mbb{Q}(i,\sqrt{5})/\mbb{Q}(i),\sigma,i) = \left\{ x_0 + x_1 \msf{e}| x_0, x_1\in\mbb{Q}(i,\sqrt{5}) \right\},
\end{equation}
where $\msf{e}^2=i$ and $z\msf{e}=\msf{e}\sigma(z)$. For the purpose of shaping, we further multiply the signal with $\alpha = 1+i\sigma(\theta)$, where $\theta=\frac{1+\sqrt{5}}{2}$ and $\bar{\theta} \defeq 1-\theta = \sigma(\theta)$. The golden code (restricted to the maximal order $\bar{\mc{A}}$ of $\mc{A}$) is then given by
\begin{align}\label{eqn:GC}
    &\mc{G} = \left\{\left.\frac{1}{\sqrt{5}}\begin{pmatrix}
                      \alpha x_0  & \alpha x_1 \\
                      i\sigma(\alpha x_1)  & \sigma(\alpha x_0) \\
                    \end{pmatrix} \right| x_0,x_1\in\mbb{Z}[i][\theta]
    \right\} \nonumber \\
    &= \left\{\left.\frac{1}{\sqrt{5}}\begin{pmatrix}
                      \alpha (a+b\theta) & \alpha (c+d\theta) \\
                      i\sigma(\alpha) (c+d\bar{\theta}) & \sigma(\alpha)(a+b\bar{\theta}) \\
                    \end{pmatrix} \right|a,b,c,d\in\Zi
    \right\}.
\end{align}
For any $A= (a+b\theta) + (c+d\theta)\msf{e}\in\bar{\mc{A}}$, we define vec$(A)\defeq(a,b,c,d)^T$ a 4-dimensional $\Zi$ vector representation with respect to the basis $\{1,\theta,\msf{e},\theta\msf{e}\}$. We see that vec$(BA)$ is a $\mathbb{Z}[i]$ lattice with generator matrix ${\bf M}(A)$. It can be verified that vec$(BA)=\mathbf{M}(A)$vec$(B)$ where
\begin{equation}\label{eqn:vec_fx}
    \mathbf{M}(A) = \begin{pmatrix}
                      a & b & i(c+d) & -id \\
                      b & a+b & -id & ic \\
                      c & d & a+b & -b \\
                      d & c+d & -b & a \\
                    \end{pmatrix},
\end{equation}
and $\det(\mathbf{M}(A))= N_{\text{rd}}(A)^2$ where
\begin{equation}
    N_{\text{rd}}(A)= \det\left[\begin{pmatrix}
      a+b\theta & c+d\theta \\
      i(c+d\bar{\theta}) & a+b\bar{\theta} \\
    \end{pmatrix}\right],
\end{equation}
is the reduced norm of $A$.

\subsection{Golden-coded index coding}
In \cite{viterbo05partitionGC} and \cite{hong07GC}, a one-to-one mapping $\Psi:\mbb{K}\rightarrow \mc{A}$ between elements in $\mc{A}$ and elements in $\mbb{K} = \mbb{Q}(\msf{e},\sqrt{5})$ was defined. Such mapping is a group homomorphism between the additive groups of $\mc{A}$ and of $\mbb{K}$. In order to find the appropriate subcodes of $\mc{G}$ we partition $\Ok$ (the ring of integers of $\mbb{K}$), which in turn will give us a partition of $\bar{\mc{A}}$. The corresponding golden codewords are obtained by writing the elements of $\bar{\mc{A}}$ in the matrix form. This method has been adopted in \cite{viterbo05partitionGC, hong07GC} for partitioning $\mc{G}$ into golden subcodes for golden space-time trellis coded modulation. Unfortunately, the mapping $\Psi$ is not a group homomorphism between the multiplicative groups of $\mbb{K}$ and $\mc{A}$, since one is commutative and the other is not.
For this reason the partitions through $\mbb{K}$ may not always lead to corresponding subcodes with the desired reduced norm. Specifically, let $\phi$ be an element in $\Ok$ with the corresponding $\Psi(\phi) \in \bar{A}$. Given a principal (two-sided) ideal $\phi \Ok$, then $\Psi(\phi \Ok)$ and the principal left ideal $\bar{\mc{A}}\Psi(\phi)$ are in general different. In general, nothing can be said about the reduced norm of the elements of $\bar{\mc{A}} \Psi(\phi)$ from $\Psi(\phi\Ok)$. The following lemma establishes some cases where these two are the same.
\begin{lemma}\label{lma:two_sided}
    If $\phi=\alpha + \beta \msf{e}$ where $\alpha, \beta\in\mbb{Z}[i]$, then $\Psi(\phi \Ok)= \bar{\mc{A}}\Psi(\phi)$.
\end{lemma}
\begin{IEEEproof}
    For every $A=(a + b\theta) + (c + d\theta)\msf{e}\in\Ok$, $\phi A = [\alpha(a+b\theta) + \beta(c+d\theta)] + [\beta(a+b\theta) + \alpha(c+d\theta)]\msf{e}$. This is exactly what we obtain if we compute $\Psi(A) \cdot \Psi(\phi)$. Thus $\Psi(\phi \Ok)=\bar{\mc{A}}\Psi(\phi)$.
\end{IEEEproof}

From this point forward, we abuse the notation by using the same $\phi$ to denote $\phi\in\Ok$ and $\Psi(\phi)\in\bar{\mc{A}}$.
Let $\phi_1,\ldots,\phi_K$ be elements of the form $\alpha+\beta\msf{e}$ in $\Ok$ that are relatively prime to each other. i.e., $\phi_k\Ok +\phi_l\Ok = \Ok$ for $k\neq l \in \{1,\ldots,K\}$. Let $q=\phi_1\cdot\ldots\cdot\phi_K$ and let $q_k=\phi_1\cdot\ldots\phi_{k-1}\phi_{k+1}\ldots\cdot\phi_K$. We thus have the following partition $\Ok = q_1 \Ok + \ldots + q_K\Ok$ and
\begin{equation}
    \Ok/q\Ok = q_1 \Ok/q\Ok \oplus \ldots \oplus q_K\Ok/q\Ok,
\end{equation}
where the direct sums are guaranteed by the Chinese remainder theorem \cite[Corollary 2.27]{Hungerford74}. Now Lemma~\ref{lma:two_sided} implies that
\begin{equation}\label{eqn:partition_GA}
    \bar{\mc{A}}/\bar{\mc{A}}q = \bar{\mc{A}} q_1/\bar{\mc{A}}q \oplus \ldots \oplus \bar{\mc{A}}q_K / \bar{\mc{A}} q.
\end{equation}

For each $k\in\{1,\ldots,K\}$, we can represent $\bar{\mc{A}}q_k$ via \eqref{eqn:vec_fx} to get a $\Zi$-lattice $\Lambda_k=$vec$(\bar{\mc{A}} q_k)$ with a generator matrix
\begin{equation}\label{eqn:G_k}
    \mathbf{G}_k = \mathbf{M}(q_k)\mathbf{G},
\end{equation}
where $\mathbf{G}$ is a generator matrix of the base lattice $\Lambda=$vec$(\bar{\mc{A}})$. Also, we let $\Lambda_s=\text{vec}(\bar{\mc{A}}q)$. It is clear that $\Lambda_s\subset\Lambda_k\subset\Lambda$ and the order of the coset decomposition is given by
\begin{align}
    |\bar{\mc{A}} q_k/\bar{\mc{A}} q|&=|\Lambda_k/\Lambda_s| = \frac{|\det(\mathbf{M}(q)\mathbf{G})|^2}{|\det(\mathbf{M}(q_k)\mathbf{G})|^2} \nonumber \\
    &= \frac{|\det(\mathbf{M}(q))\det(\mathbf{G})|^2}{|\det(\mathbf{M}(q_k))\det(\mathbf{G})|^2} \nonumber \\
    &= \frac{|\det(\mathbf{M}(q))|^2}{|\det(\mathbf{M}(q_k))|^2} 
    = |N_{\text{rd}}(\phi_k)|^4.
\end{align}
The following lemma further establishes the relationship between the lattice partition and the coset decomposition of $\bar{\mc{A}}$.
\begin{lemma}\label{lma:comp_coset_A}
    $\Lambda\hspace{-3pt}\mod\Lambda_s$ corresponds to a complete set of coset leader of the quotient algebra $\bar{\mc{A}}/\bar{\mc{A}} q$.
\end{lemma}
\begin{IEEEproof}
    Let $\boldsymbol\lambda_1, \boldsymbol\lambda_2 \in\Lambda$ such that $\boldsymbol\lambda_1 =\text{vec}(g_1)$ and $\boldsymbol\lambda_2 =\text{vec}(g_2)$ where $g_1,g_2\in\bar{\mc{A}}$. Moreover, let us assume $\boldsymbol\lambda_1 \equiv \boldsymbol\lambda_2\hspace{-3pt}\mod\Lambda_s$. We have
    \begin{align}
        &\hphantom{(\Leftrightarrow)}~ \boldsymbol\lambda_1 - \boldsymbol\lambda_2 \equiv\mathbf{0}\hspace{-3pt}\mod\Lambda_s \nonumber \\
        &(\Leftrightarrow)~ \text{vec}(g_1) - \text{vec}(g_2) \equiv\mathbf{0}\hspace{-3pt}\mod\Lambda_s \nonumber \\
        &(\Leftrightarrow)~ \text{vec}(g_1-g_2) \equiv\mathbf{0}\hspace{-3pt}\mod\Lambda_s \nonumber \\
        &(\Leftrightarrow)~ \text{vec}(g_1 - g_2) \in \Lambda_s.
    \end{align}
    Moreover, since the vectorization operation is bijective, we have $g_1-g_2 \in \bar{\mc{A}} q$ which results in $g_1\equiv g_2\hspace{-3pt}\mod \bar{\mc{A}} q$. We conclude the proof by noting that $|\Lambda/\Lambda_s| = |\bar{\mc{A}}/\bar{\mc{A}} q|$.
\end{IEEEproof}
\begin{remark}
    In what follows, we would like to construct golden-coded index coding based on the partition of the cyclic division algebra $\bar{\mc{A}}\hspace{-3pt}\mod \bar{\mc{A}}q$ where the modulo is based on a division algorithm that yields a remainder with a smaller reduced norm than the one of the divisor. This only guarantees that the overall codebook would have the minimum reduced norm but in general, does not guarantee the minimum Euclidean norm. Consequently, the code could have a very bad shape and may result in a significant shaping loss. Fortunately, the above lemma has guaranteed the one-to-one mapping between $\bar{\mc{A}}\hspace{-3pt}\mod \bar{\mc{A}}q$ and $\Lambda\hspace{-3pt}\mod \Lambda_s$ and hence our construction will be based on $\Lambda\hspace{-3pt}\mod \Lambda_s$, which automatically takes care of shaping.
\end{remark}

The proposed golden-coded index coding exploits the partition in \eqref{eqn:partition_GA}. Specifically, we set
\begin{equation}\label{eqn:rate_Wk}
    W_k = |\bar{\mc{A}} q_k/\bar{\mc{A}} q| =|N_{\text{rd}}(\phi_k)|^4,
\end{equation}
and generate individual constellation $\Lambda_k\hspace{-3pt}\mod \Lambda_s$. We then use an arbitrary bijective mapping $\varphi_k$ to map each $w_k$ to $\mathbf{x}_k=\varphi_k(w_k)\in\Lambda_k\hspace{-3pt}\mod \Lambda_s$ and form 
\begin{equation}
    \mathbf{x}=\left(\mathbf{x}_1+\ldots+\mathbf{x}_K\right)\hspace{-3pt}\mod\Lambda_s.
\end{equation}
Note that from Lemma~\ref{lma:comp_coset_A} and the partition in \eqref{eqn:partition_GA},
\begin{equation}
    \mathbf{x}\in\sum_{k=1}^K \Lambda_k/\Lambda_s\hspace{-3pt}\mod\Lambda_s = \Lambda \hspace{-3pt}\mod\Lambda_s.
\end{equation}

Note that $\Lambda_k$ and $\Lambda$ are 4-dimensional $\Zi$ lattices; thus, $\mathbf{x}=(a,b,c,d)^T$ for some $a,b,c,d\in\Zi$. We then form the proposed golden-coded index coding as
\begin{align}
      \mc{C} =& \left\{ \left.\frac{1}{\sqrt{5}}\begin{pmatrix}
                      \alpha (a+b\theta)  & \alpha (c+d\theta) \\
                      i\sigma(\alpha) (c+d\bar{\theta})  & \sigma(\alpha)(a+b\bar{\theta}) \\
                    \end{pmatrix} \right|\right. \nonumber \\
                    &\hspace{2.7cm}\left. \vphantom{\frac{1}{\sqrt{5}}} (a,b,c,d)^T\in \Lambda\hspace{-3pt}\mod\Lambda_s
    \right\}.
\end{align}

Equipped with all the individual encoders $\varphi_k$, the $l$-th receiver first forms $\mathbf{x}_k=\varphi_k(w_k)$ for $k\in\mc{S}_l$. It then uses lattice decoding to decode the received signal to the nearest element in the Golden subcode corresponding to
\begin{equation}
    \left(\sum_{k\in\mc{S}_l} \mathbf{x}_k + \sum_{k'\notin\mc{S}_l}\Lambda_{k'}/\Lambda_s\right)\hspace{-3pt}\mod\Lambda_s.
\end{equation}
The Golden subcode at the $l$-th receiver becomes a coset of
\begin{align}\label{eqn:C_sl}
      \mc{C}_{\mc{S}_l} =& \left\{ \vphantom{\sum_{k\notin\mc{S}_l}} \left.\frac{1}{\sqrt{5}}\begin{pmatrix}
                      \alpha (a+b\theta)  & \alpha (c+d\theta) \\
                      i\sigma(\alpha) (c+d\bar{\theta})  & \sigma(\alpha)(a+b\bar{\theta}) \\
                    \end{pmatrix} \right|\right. \nonumber \\
                    &\hspace{1.2cm}\left.  (a,b,c,d)^T\in \sum_{k'\notin\mc{S}_l}\Lambda_{k'}/\Lambda_s\hspace{-3pt}\mod\Lambda_s
    \right\}.
\end{align}

We now show the main result of this section.
\begin{theorem}\label{thm:GC_unif_SI}
    For any $\mc{S}_l\subset\{1,\ldots,K\}$, the proposed golden-coded index coding provides uniform side information gain of $6$ dB.
\end{theorem}
\begin{IEEEproof}
    From \eqref{eqn:rate_Wk}, $w_{\mc{S}_l}$ has a rate
\begin{equation}\label{eqn:GC_Rsl}
    R_{\mc{S}_l} = \frac{1}{8}\sum_{k\in\mc{S}_l}\log_2 |N_{\text{rd}}(\phi_k)|^4\quad\text{bits/real dimension.}
\end{equation}
We note that shifting by a constant will not change the lattice structure; therefore, we henceforth assume $w_k=0$ for every $k\in\mc{S}_l$. From \eqref{eqn:C_sl}, after revealing $w_{\mc{S}_l}$, each $\mathbf{X}\in\mc{C}_{\mc{S}_l}$ corresponds to $(a,b,c,d)^T\in\sum_{k\notin\mc{S}_l}\Lambda_{k}/\Lambda_s\hspace{-3pt}\mod\Lambda_s$ or equivalently an element $x_0+x_1\msf{e}\in\sum_{k\notin\mc{S}_l}\bar{\mc{A}} q_k/\bar{\mc{A}} q$ where $x_0=a+b\theta$ and $x_1=c+d\theta$. Let $\eta_l$ be a generator of the left ideal $\sum_{k\notin\mc{S}_l}\bar{\mc{A}}q_k$. From Lemma~\ref{lma:two_sided}, $\sum_{k\notin\mc{S}_l}\bar{\mc{A}}q_k=\Psi\left(\sum_{k\notin\mc{S}_l}q_k\Ok\right)$. Hence, $\eta_l$ is also a generator of $\sum_{k\notin\mc{S}_l}q_k\Ok$ and therefore $\eta_l$ and $\prod_{k\in\mc{S}_l}\phi_k$ are associates. Without loss of generality, we set $\eta_l=\prod_{k\in\mc{S}_l}\phi_k$.

The determinant of $\mathbf{X}\in\mc{C}_{\mc{S}_l}$ can be computed as follows,
\begin{align}
    \det(\mathbf{X}) &= \frac{1}{5}N_{\text{rd}}(\alpha) \det\left[\begin{pmatrix}
                                             a+b\theta & c+d\theta \\
                                             i(c+d\bar{\theta}) & a+b\bar{\theta} \\
                                           \end{pmatrix}
    \right] \nonumber \\
    &= \frac{1}{5}N_{\text{rd}}(\alpha)N_{\text{rd}}(x_0+x_1\msf{e}) \geq \frac{1}{5}N_{\text{rd}}(\alpha)N_{\text{rd}}(\eta_l),
\end{align}
where the last inequality is due to the fact that the reduced norm is multiplicative and $\eta_l$ is a generator of $\sum_{k\notin\mc{S}_l}q_k\Ok$. Therefore, plugging in $|N_{\text{rd}}(\alpha)|^2=5$ results in
\begin{align}\label{eqn:GC_delta_l}
    \delta_l 
    &= \frac{1}{5}|N_{\text{rd}}(\eta_l)|^2 = \frac{1}{5}\prod_{k\in\mc{S}_l} |N_{\text{rd}}(\phi_k)|^2.
\end{align}
Combining \eqref{eqn:GC_Rsl}, \eqref{eqn:GC_delta_l}, and the fact that $\delta = N(1)^2/5 = 1/5$ results in
\begin{align}
    \Gamma(\mc{C},\mc{S}_l) 
    &= \frac{20\sum_{k\in\mc{S}_l}\log_{10}|N_{\text{rd}}(\phi_k)|^2}{\sum_{k\in\mc{S}_l}\log_2 |N_{\text{rd}}(\phi_k)|^2}\approx 6~\text{dB}.
\end{align}
\end{IEEEproof}

\section{Simulation Results}\label{sec:simu_result}
We now provide some examples and simulation results. We first use {\sc Magma} \cite{magma} to tailor numbers into primes in $\Ok$. We specifically look for elements of the form in Lemma~\ref{lma:two_sided}. Some examples are given below.
\begin{example}
    We have the partition of the principal ideal $2 \Ok = \left((1+i\msf{e})\Ok\right)^4$ and $N_{\text{rd}}(1+i\msf{e}) = 1+ i$. This partition has been adopted to construct golden space-time trellis coded modulation in \cite{viterbo05partitionGC, hong07GC}.
\end{example}
\begin{example}\label{ex:p17}
    The principal ideal $17\Ok$ has the partition $17\Ok = \mc{I}_1\cdot\mc{I}_2\cdot\mc{I}_3\cdot\mc{I}_4$ where $\mc{I}_1 = (1+2\msf{e})\Ok$, $\mc{I}_2 = (2-\msf{e})\Ok$, $\mc{I}_3 = (-i+2i\msf{e})\Ok$, and $\mc{I}_4 = (1-2i\msf{e})\Ok$. The generators of these prime ideals have the reduced norms $N_{\text{rd}}(\mc{I}_1) = 1 - 4i$, $N_{\text{rd}}(\mc{I}_2) = 4-i$, $N_{\text{rd}}(\mc{I}_3) = -1 + 4i$, and $N_{\text{rd}}(\mc{I}_4) = 1 + 4i$.
\end{example}
\begin{example}
    The principal ideal $73\Ok$ has the partition $73\Ok = \mc{I}_1\cdot\mc{I}_2\cdot\mc{I}_3\cdot\mc{I}_4$ where $\mc{I}_1 = (-2i+(i-2)\msf{e})\Ok$, $\mc{I}_2 = (2i+(i-2)\msf{e})\Ok$, $\mc{I}_3 = (1-2i-2\msf{e})\Ok$, and $\mc{I}_4 = (2-(i+2)\msf{e})\Ok$. 
\end{example}

\begin{figure}
    \centering
    \includegraphics[width=3.5in]{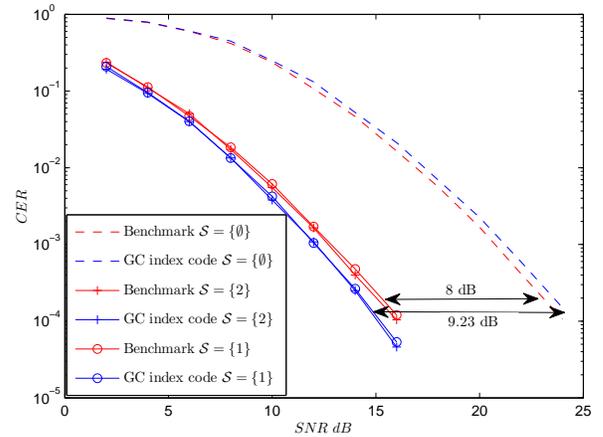}
    \caption{SNR versus CER over the MIMO Rayleigh fading network.}
    \label{fig:CER_p17}
\end{figure}

In Fig.~\ref{fig:CER_p17}, we consider $K=2$ and show codeword error rates (CER) of the proposed golden-coded index coding with $\phi_1=1+2\msf{e}$ and $\phi_2=2-\msf{e}$ in Example~\ref{ex:p17}. As a benchmark, we also partition the golden code with 16-QAM into two subcodes using the partition of 16-QAM constellation obtained in \cite{viterbo15_index_QAM}. Specifically, we use the partition in \cite[Example 2]{viterbo15_index_QAM} to partition $\mc{M}$ 16-QAM into two constellations $\mc{M}_1$ and $\mc{M}_2$, each has 8 elements. We then set $W_1=W_2=8$ and use $\mc{M}_1$ and $\mc{M}_2$ to encode $w_1$ and $w_2$, respectively. The overall code is given by
\begin{align}\label{eqn:partition_QAM}
      \hspace{-15pt}\left\{ \left.\frac{1}{\sqrt{5}}\begin{pmatrix}
                      \alpha (a+b\theta)  & \alpha (c+d\theta) \\
                      i\sigma(\alpha) (c+d\bar{\theta})  & \sigma(\alpha)(a+b\bar{\theta}) \\
                    \end{pmatrix} \right| a,b,c,d\in \mc{M}
    \right\},
\end{align}
and when $w_2$ (similarly $w_1$) is given, the code becomes \eqref{eqn:partition_QAM} with $\mc{M}$ replaced by $\mc{M}_1$ ($\mc{M}_2$).
In Fig.~\ref{fig:CER_p17}, for the proposed scheme, one observes a 9.23 dB SNR gain when either $w_1$ or $w_2$ is revealed. By inspecting the code, we obtain $N_{\mc{C}}=1872$ and $N_{\mc{C}_1}=N_{\mc{C}_2}=112$, which accounts for 3.06 dB SNR gain predicted in \eqref{eqn:snr_gain_exact} from reduction of the multiplicity of the elements having the minimum determinant. The remaining 6.17 dB gain can be predicted by the increase of minimum determinant $|N_{\text{rd}}(1+2\msf{e})|^2=|N_{\text{rd}}(2-\msf{e})|^2=17$ and results in approximate 6 dB side information gain after normalization by the rate $1.022$ bits/real dimension. One can also use \eqref{eqn:snr_gain_exact} to explain the 8 dB SNR gain observed in this figure for the golden code with QAM partition where $N_{\mc{C}}=1400$ and $N_{\mc{C}_1}=N_{\mc{C}_2}=3.75$ and the increase in the minimum determinant is 2.

Some interesting observations are as follows. We first note that the two schemes in Fig.~\ref{fig:CER_p17} have roughly the same rate and it is shown that the proposed scheme can better exploit side information (even after normalization by the respective rates). Also, the proposed scheme makes use of the algebraic structure of the golden algebra and thus has a systematic procedure while the one with QAM partition is obtained from computer simulation. Last but not least, one also observes that the side information gain of the proposed scheme largely comes from improvement of the minimum determinant, while that in the QAM partition mainly comes from reduction of the number of elements having minimum determinant. This phenomenon is quite interesting and deserves further investigation.

\section{Concluding Remarks}\label{sec:conclude}
We have partitioned the golden code into golden subcodes for the $2\times 2$ MIMO physical-layer index coding problems and successfully proposed golden-coded index coding. The partition of golden codes was based on the partition of the corresponding golden algebra, which was enabled by viewing the maximal order of it as the ring of integers of a number field. We have shown the uniform side information gain property of the proposed scheme. Simulation results have also confirmed our findings. After this, a natural next step would be to develop a general algebraic framework for partitioning other lattice space-time codes.

\bibliographystyle{IEEEtran}
\bibliography{journal_abbr,bib_lattice}

\begin{thebibliography}{10}
\providecommand{\url}[1]{#1}
\csname url@samestyle\endcsname
\providecommand{\newblock}{\relax}
\providecommand{\bibinfo}[2]{#2}
\providecommand{\BIBentrySTDinterwordspacing}{\spaceskip=0pt\relax}
\providecommand{\BIBentryALTinterwordstretchfactor}{4}
\providecommand{\BIBentryALTinterwordspacing}{\spaceskip=\fontdimen2\font plus
\BIBentryALTinterwordstretchfactor\fontdimen3\font minus
  \fontdimen4\font\relax}
\providecommand{\BIBforeignlanguage}[2]{{%
\expandafter\ifx\csname l@#1\endcsname\relax
\typeout{** WARNING: IEEEtran.bst: No hyphenation pattern has been}%
\typeout{** loaded for the language `#1'. Using the pattern for}%
\typeout{** the default language instead.}%
\else
\language=\csname l@#1\endcsname
\fi
#2}}
\providecommand{\BIBdecl}{\relax}
\BIBdecl

\bibitem{niesen14}
M.~A. Maddah-Ali and U.~Niesen, ``Fundamental limits of caching,'' \emph{IEEE
  Trans. Inf. Theory}, vol.~60, no.~5, pp. 2856--2867, May 2014.

\bibitem{ji16}
M.~Ji and G.~Caire, ``Fundamental limits of caching in wireless {D2D}
  networks,'' \emph{IEEE Trans. Inf. Theory}, vol.~62, no.~2, pp. 849--869,
  Feb. 2016.

\bibitem{paschos16}
G.~Paschos, E.~Bastug, I.~Land, G.~Caire, and M.~Debbah, ``Wireless caching:
  Technical misconceptions and business barriers,'' \emph{IEEE Commun. Mag.},
  vol.~54, no.~8, pp. 16--22, August 2016.

\bibitem{bar-yossef11}
Z.~Bar-Yossef, Y.~Birk, T.~S. Jayram, and T.~Kol, ``Index coding with side
  information,'' \emph{IEEE Trans. Inf. Theory}, vol.~57, no.~3, pp.
  1479--1494, Mar. 2011.

\bibitem{viterbo14index}
L.~Natarajan, Y.~Hong, and E.~Viterbo, ``Lattice index coding,'' \emph{IEEE
  Trans. Inf. Theory}, vol.~61, no.~12, pp. 6505--6525, Dec. 2015.

\bibitem{huang15_lic}
Y.-C. Huang, ``Lattice index codes from algebraic number fields,'' \emph{IEEE
  Trans. Inf. Theory}, vol.~63, no.~4, pp. 2098--2112, Apr. 2017.

\bibitem{oggier07}
F.~Oggier, J.-C. Belfiore, and E.~Viterbo, ``Cyclic division algebras: A tool
  for space-time coding,'' \emph{Foundations and Trends in Communications and
  Information Theory}, vol.~4, no.~1, pp. 1--95, 2007.

\bibitem{belfiore05GC}
J.-C. Belfiore, G.~Rekaya, and E.~Viterbo, ``The golden code: A $2\times 2$
  full-rate space-time code with nonvanishing determinants,'' \emph{IEEE Trans.
  Inf. Theory}, vol.~51, no.~4, pp. 1432--1436, Apr. 2005.

\bibitem{viterbo05partitionGC}
D.~Champion, J.-C. Belfiore, G.~Rekaya, and E.~Viterbo, ``Partitioning the
  {Golden} code: A framework to the design of space-time coded modulation,'' in
  \emph{Proc. Canadian Workshop on Inf. Theory}, Jun. 2005.

\bibitem{hong07GC}
Y.~Hong, E.~Viterbo, and J.-C. Belfiore, ``Golden space-time trellis coded
  modulation,'' \emph{IEEE Trans. Inf. Theory}, vol.~53, no.~5, pp. 1689--1705,
  May 2007.

\bibitem{Hungerford74}
T.~W. Hungerford, \emph{Algebra (Graduate Texts in Mathematics)}.\hskip 1em
  plus 0.5em minus 0.4em\relax Springer, 1974.

\bibitem{magma}
\BIBentryALTinterwordspacing
W.~Bosma, J.~Cannon, and C.~Playoust, ``The {M}agma algebra system. {I}. {T}he
  user language,'' \emph{J. Symbolic Comput.}, vol.~24, no. 3-4, pp. 235--265,
  1997. [Online]. Available: \url{http://dx.doi.org/10.1006/jsco.1996.0125}
\BIBentrySTDinterwordspacing

\bibitem{viterbo15_index_QAM}
L.~Natarajan, Y.~Hong, and E.~Viterbo, ``Index codes for the {Gaussian}
  broadcast channel using quadrature amplitude modulation,'' \emph{IEEE Commun.
  Lett.}, vol.~19, no.~8, pp. 1291--1294, Aug. 2015.

\end{thebibliography}

\end{document}